\documentclass[12pt]{article}
\usepackage{epsf}
\usepackage{graphicx}
\usepackage{axodraw}
\usepackage{amssymb}
\usepackage{cite}

\begin{document}
\vskip -1.5cm
\begin{flushright}
hep-ph/0508077\\
August 2005
\end{flushright}

\begin{center}
{\LARGE {\bf Heavy Majorana Neutrino Production }}\\[0.3cm]
{\LARGE {\bf at {\boldmath $e^-\gamma$} Colliders}}\\[1.4cm]
{\large Simon Bray, Jae Sik Lee and Apostolos Pilaftsis }\\[0.3cm]
{\em Department of Physics and Astronomy, University of Manchester,\\
Manchester M13 9PL, United Kingdom}
\end{center}
\vskip0.7cm \centerline{\bf  ABSTRACT} {\small We  study signatures of
heavy  Majorana  neutrinos  at  $e^-\gamma$  colliders.   Since  these
particles violate lepton  number, they can give rise  to the reactions
$e^-\gamma\rightarrow     W^-W^-e^+$     and     $e^-\gamma\rightarrow
W^+\mu^-\mu^-\nu$. The Standard  Model background contains extra light
neutrinos that escape detection, and can be reduced to an unobservable
level  after imposing  appropriate kinematical  cuts.  We  analyze the
physics  potential of  an $e^-\gamma$  collider, with  centre  of mass
energies  $\sqrt{s}=0.5$--1~TeV and a  total integrated  luminosity of
100~fb$^{-1}$,  for  detecting heavy  Majorana  neutrinos with  masses
$m_N=100$--400~GeV.     Assuming   that    heavy    neutrinos   couple
predominantly to  only one lepton flavour  at a time, we  find that 10
expected   background-free   events   can   be   established   for   a
$WeN$-coupling  $|B_{eN}| \stackrel{>}{{}_\sim} 4.6\times  10^{-3}$ or
for a  $W\mu N$-coupling $|B_{\mu  N}| \stackrel{>}{{}_\sim} 9.0\times
10^{-2}$.  Instead,  if no  signal is observed,  this will  imply that
$|B_{eN}|  \stackrel{<}{{}_\sim} 2.7\times  10^{-3}$ and  $|B_{\mu N}|
\stackrel{<}{{}_\sim} 0.05$ at the 90\% confidence level.}

\bigskip

\section{Introduction}

Neutrino   oscillation  experiments  \cite{Fukuda:2002pe,Ahmed:2003kj,
Eguchi:2002dm}  have  established the  fact  that  the observed  light
neutrinos  are not  strictly massless,  as predicted  in  the Standard
Model~(SM),  but  have   non-zero  tiny  masses.   One  well-motivated
framework for giving small masses to  neutrinos is to extend the SM by
adding one right-handed  neutrino per family. In such  a scenario, the
right-handed neutrinos are  singlets under the SM gauge  group, and so
are allowed  to have large  Majorana masses in the  Lagrangian.  Apart
from  the  three  observable  neutrinos $\nu_{1,2,3}$,  the  resulting
spectrum  will contain three  additional heavy  neutrinos $N_{1,2,3}$,
whose masses  could range  from the electroweak  to the  Grand Unified
Theory  (GUT) scale~\cite{Minkowski:1977sc}.   Direct searches  at LEP
put  strong  limits  on  the  couplings of  heavy  Majorana  neutrinos
weighing  less than  about  100~GeV~\cite{Achard:2001qv}.  For  larger
heavy Majorana  masses, the indirect limits are  still significant and
usually arise from the  non-observation of sizeable quantum effects on
low-energy   observables,   as   well   as   from   the   absence   of
lepton-flavour-violating
decays~\cite{Langacker:1988ur,Korner:1992an,Burgess:1993vc,Nardi:1994iv,
Cheng:1980tp,Ilakovac:1994kj,Bergmann:1998rg,Illana:2000ic,Cvetic:2002jy},
e.g.~$\mu \to  e\gamma$, $\mu \to eee$  and $\mu \to  e$ conversion in
nuclei etc.

In this  paper we  study the distinctive  signatures of  lepton number
violation  (LNV) at  an $e^-\gamma$  collider, which  are  mediated by
heavy  Majorana  neutrinos.   In   particular,  we  consider  the  LNV
reactions  $e^-\gamma\rightarrow W^-W^-e^+$  and $e^-\gamma\rightarrow
W^+\mu^-\mu^-\nu$.  These processes are  strictly forbidden in the SM,
and the corresponding background  will always involve additional light
neutrinos.  An  $e^-\gamma$ collider  provides a clean  environment to
look for  such particles,  since the initial  state $e^-\gamma$  has a
definite non-zero lepton number and hence any LNV signal can easily be
detected.  There are already realistic proposals~\cite{Badelek:2001xb}
for  the construction  of an  $e^-\gamma$ machine,  which will  run at
0.5--1~TeV centre of mass system (CMS) energies and a total integrated
luminosity  of~100~fb$^{-1}$.  Here,   we  will  analyze  the  physics
potential of such a machine  to discover heavy Majorana neutrinos with
masses between 100 and 400~GeV.

Our study has been structured as follows.  After briefly reviewing the
SM with right  handed neutrinos in Section~2, we  discuss in Section~3
the LNV  signals due  to heavy Majorana  neutrinos and  the associated
backgrounds.  In Section~4 we  present numerical results for the cross
sections of the reactions $e^-\gamma \to W^-W^-e^+$ and $e^-\gamma \to
W^+\mu^-\mu^-\nu$.   In   particular,  taking  into   account  the  SM
backgrounds, we place lower limits  on the heavy neutrino couplings to
the  charged leptons  and the  $W^\pm$  bosons, by  requiring that  an
observable  LNV  signal is  obtained.   Finally,  our conclusions  are
summarized in Section 5.

\setcounter{equation}{0}
\section{The Standard Model with Right-Handed Neutrinos}

We briefly review the relevant low-energy structure of the SM modified
by the  presence of right-handed  neutrinos. The Lagrangian describing
the neutrino masses and mixings reads:
\begin{equation}
\mathcal{L}^{\rm mass}_{\nu}\ =\ -\frac{1}{2}\left(\begin{array}{cc}
\overline{\nu_L^0} & \overline{{(\nu_R^0)}^c}
\end{array}\right)
\left(\begin{array}{cc}
0 & m_D \\ m_D^T & M_R
\end{array}\right)
\left(\begin{array}{c}
{(\nu_L^0)}^c \\ \nu_R^0
\end{array}\right)\ +\ {\rm h.c.},
\end{equation}
where  $\nu_L^0 =  ( \nu_{e  L},\ \nu_{\mu  L},\ \nu_{\tau  L})^T$ and
$\nu_R^0 = ( \nu_{e R},\  \nu_{\mu R},\ \nu_{\tau R} )^T$ collectively
denote the left-  and right-handed neutrino fields in  the weak basis,
and $m_D$  and $M_R$  are $3 \times  3$ complex matrices.   The latter
obeys  the Majorana  constraint $M_R  =  M^T_R$ and,  without loss  of
generality,  can be  assumed to  be  diagonal and  positive. The  weak
neutrino states are related to the Majorana mass eigenstates through:
\begin{equation}
\left(\begin{array}{cc}\nu_L \\ N_L \end{array}\right)\ =\ U^T
\left(\begin{array}{c}\nu_L^0 \\ {(\nu_R^0)}^c\end{array}\right)\;,
\end{equation}
where $U$ is a $6 \times 6$ unitary matrix.

In  typical seesaw  scenarios~\cite{Minkowski:1977sc}, the  Dirac mass
terms $m_D$ are expected to be around the electroweak scale, e.g.~$m_D
\sim M_l$ or $m_D \sim M_u$, where $M_l$ ($M_u$) is the charged lepton
(up-quark) mass  matrix, whilst the Majorana mass  $M_R$ being singlet
under the  SM gauge group may be  very large, close to  the GUT scale.
Seesaw models can explain the smallness of the observed light neutrino
masses, generically predicted to be $\sim m^2_D/M_R$, through the huge
hierarchy  between $m_D$  and $M_R$.   However, the  couplings  of the
heavy neutrinos to SM  particles will be generically highly suppressed
$\sim  m_D/M_R$,  thus  rendering  the  direct  observation  of  heavy
Majorana neutrinos impossible.

Another and perhaps phenomenologically  more appealing solution to the
smallness of  the light neutrino  masses may arise due  to approximate
flavour  symmetries  that  may  govern  the Dirac  and  Majorana  mass
matrices $m_D$ and $M_R$~\cite{EWitten,Pilaftsis:1991ug,Gluza:2002vs}.
In such  models, the heavy-to-light Majorana couplings  are also $\sim
m_D/M_R$, but they  can be completely unrelated to  the light neutrino
mass matrix ${\bf m}_\nu$, which  is determined by the relation: ${\bf
m}_\nu =  - m_D M^{-1}_R m^T_D$.   In such non-seesaw  models, one may
have  $m_D  \sim  M_l$  and  $M_R  \sim  100$~GeV,  without  being  in
contradiction with neutrino data. However, as we will see below, these
models  are  constrained  by  electroweak  precision  data  and  other
low-energy  lepton-flavour/number-violating observables. Our  study of
LNV signatures at an $e^-\gamma$ collider will focus on such non-seesaw
realizations.

For our subsequent phenomenological discussion,   we now exhibit   the
interaction  Lagrangians  of the heavy neutrinos   to $W^\pm$, $Z$ and
Higgs ($H$) bosons~\cite{Pilaftsis:1991ug}:
\begin{eqnarray}
\mathcal{L}_W & = & -\,\frac{g}{\sqrt{2}}\,W_\mu^-\, \overline{l}_l
\gamma^\mu P_L B_{l j} \left(\begin{array}{c} \nu \\ N
\end{array}\right)_j\ +\ {\rm h.c.},\\ 
\mathcal{L}_Z & = &
-\,\frac{g}{4c_w}Z_\mu\; \left(\begin{array}{cc} \overline{\nu} &
\overline{N}\end{array}\right)_i\gamma^\mu \Big(\,i{\rm Im}\,C_{ij}\: -\:
\gamma_5\, {\rm Re}\,C_{ij}\,\Big)\,\left(\begin{array}{c}\nu \\
N\end{array}\right)_j\; ,\\ 
\mathcal{L}_H & = &
-\, \frac{g}{4M_W}H\; \left(\begin{array}{cc}\overline{\nu} &
\overline{N}\end{array}\right)_i\Big[\,(m_i + m_j)\,
{\rm Re}\,C_{ij}\nonumber \\
&&-\:i\gamma_5 (m_i-m_j)\, {\rm Im}\,C_{ij}\,\Big]\,
\left(\begin{array}{c}\nu \\
N\end{array}\right)_j\; ,
\end{eqnarray}
where $m_{i,j}$ (with  $i,j = 1,2,\dots 6$) denote  the physical light
and heavy neutrino masses, and
\begin{equation}
  \label{BC}
B_{lj} \ = \ \sum_{k=1}^3
V^L_{li} U_{kj}^*\;,\qquad C_{ij} \ = \ \sum_{k=1}^3 U_{ki}U_{kj}^*\; ,
\end{equation}
with $l = e,\mu,\tau$.  In~(\ref{BC}),  $V^L$ is a $3\times 3$ unitary
matrix  relating  the weak  to  mass  eigenstates  of the  left-handed
charged leptons.

As  was   mentioned  above,  the  mixing  elements   $B_{lj}$  can  be
constrained    from    LEP    and    low-energy    electroweak    data
\cite{Langacker:1988ur,Korner:1992an,Burgess:1993vc,Nardi:1994iv,
Cheng:1980tp,Ilakovac:1994kj,Bergmann:1998rg,Illana:2000ic,Cvetic:2002jy}.
Following \cite{delAguila:2005mf}, we define
\begin{equation}
\Omega_{ll'}\ \equiv\
\delta_{ll'}-\sum_{i=1}^3B_{l\nu_i}B_{l'\nu_i}^*\ =\ 
\sum_{i=1}^3B_{lN_i}B_{l'N_i}^*\; .
\end{equation}
The  above definition  is  a generalization  of the  Langacker--London
parameters  $(s_L^{\nu_{e,\mu,\tau}})^2$~\cite{Langacker:1988ur}, with
the  identification:  $\Omega_{ll}=(s_L^{\nu_l})^2$.   At  the  $90\%$
confidence  level   (CL),  the  allowed  values   for  the  parameters
$\Omega_{ll}$ are \cite{Bergmann:1998rg}:
\begin{equation}
\Omega_{ee}\ \le\ 0.012\,,\qquad
\Omega_{\mu\mu}\ \le\ 0.0096\,,\qquad
\Omega_{\tau\tau}\ \le\ 0.016\; .
\label{2}
\end{equation}
These limits depend only weakly on the heavy neutrino masses.

Lepton-flavour-violating            processes,            e.g.~$\mu\to
e\gamma$~\cite{Cheng:1980tp},    $\mu     \to    eee$,    $\tau    \to
eee$~\cite{Ilakovac:1994kj,Illana:2000ic,Cvetic:2002jy},     can    be
induced by  heavy neutrino quantum effects.  Such  processes do depend
on      the      heavy       neutrino      masses      and      Yukawa
couplings~\cite{Ilakovac:1994kj}.   For  $m_N \gg  M_W$  and $m_D  \ll
M_W$,   the  limits   derived   after  including   the  recent   BaBar
constraints~\cite{BaBar}   from   $B   (\tau   \to  e,\mu   \gamma   )
\stackrel{<}{{}_\sim} 10^{-7}$ are
\begin{equation}
  \label{3}
\vert\Omega_{e\mu}\vert\ \stackrel{<}{{}_\sim}\ 0.0001 \,,\qquad
\vert\Omega_{e\tau}\vert\ \stackrel{<}{{}_\sim}\ 0.02\,,\qquad
\vert\Omega_{\mu\tau}\vert\ \stackrel{<}{{}_\sim}\  0.02\; .
\end{equation}

Finally,  we present  the partial  decay  widths of  a heavy  Majorana
neutrino $N$ for its dominant decay channels~\cite{Pilaftsis:1991ug}:
\begin{eqnarray}
\Gamma(N\rightarrow l^\pm W^\mp) & = & \frac{\alpha_w{\vert
B_{l N}\vert}^2}{16\, M_W^2m_N^3}\ (m_N^2-M_W^2)^2\: (m_N^2 + 2M_W^2)\; ,
\label{4} \\ 
\Gamma(N\rightarrow \nu_i Z) & = & \frac{\alpha_w\,{\vert
C_{\nu_i N}\vert}^2}{16\, M_W^2m_N^3}\ (m_N^2-M_Z^2)^2\: (m_N^2 +
2M_Z^2)\; ,
\label{5} \\ 
\Gamma(N\rightarrow \nu_i H) & = & \frac{\alpha_w\,{\vert
C_{\nu_i N}\vert}^2}{16\, M_W^2m_N}\ (m_N^2-M_H^2)^2\; ,
\label{6}
\end{eqnarray}
where  $\alpha_w   =  g^2/(4\pi)$.   Notice  that   up  to  negligible
corrections  ${\cal O}(m^4_D/M^4_R)$,  it  can be  shown that  $\sum_l
|B_{lN}|^2  = \sum_i  |C_{\nu_i N}|^2$.   Hence, for  heavy neutrinos,
with  $m_N \gg  M_H$,  decaying  into all  charged  leptons and  light
neutrinos,  one  obtains  the  useful  relation  among  the  branching
fractions: $B (N\to l^+ W^-) = B ( N \to l^-W^+) = B (N \to \nu Z) = B
( N \to \nu H) = 1/4$.

\setcounter{equation}{0}
\section{Lepton Number Violating Signatures}

Previous analyses of single heavy neutrino production have mainly been
focused                on                $e^+e^-$               linear
colliders~\cite{BG,delAguila:2005mf,Almeida:2000yx}, where the process
$e^+e^-  \to  N\nu \to  l^{\pm}  W^{\mp}\nu$  is  considered.  Such  a
production channel,  however, has two problems not  associated with an
$e^-\gamma$  collider.   Firstly,  since  the light  neutrinos  escape
detection  with their  chirality undetermined,  the  possible Majorana
nature of  the heavy neutrinos has  very little effect  on the signal,
which makes the reduction of the contributing SM background very hard.
Secondly, an observable signal would require a heavy neutrino coupling
to  the  electron  and  the  $W^\pm$ bosons  of  reasonable  strength,
i.e.~$|B_{eN}|\stackrel{>}{{}_\sim}    10^{-2}$,    whereas   at    an
$e^-\gamma$  collider a  signal can  still  be seen  for $|B_{\mu  N}|
\stackrel{>}{{}_\sim} 0.1$, even if $B_{e N} = 0$.

Another possible  option which  might have the  same advantages  as an
$e^-\gamma$ collider  is a  $\gamma\gamma$ collider.  Here,  the heavy
Majorana neutrinos can be  produced and observed via $\gamma\gamma \to
W^+N l^- \to W^+W^+l^-l^-$~\cite{Peressutti:2002nf}. However, like the
case of the  $e^+e^-$ collider, one still has to  assess the impact of
the contributing SM background.

Depending on the  relative strength of the heavy  neutrino coupling to
the electron $B_{eN}$, we  consider the two reactions: (i)~$e^- \gamma
\to W^-  W^- l^+$ and (ii)~$e^-  \gamma \to W^+ l^-  l^- \nu$ provided
that $e^-  \gamma \to W^- W^\mp  l^\pm$ is not observed  (see also our
discussion  at   the  end  of   Section  3.3).   Both   processes  are
manifestations of LNV, and as such, they allow to eliminate major part
of the SM background.   The process~(i) dominates for relatively large
values        of       the        mixing        factor       $B_{eN}$,
i.e.~$B_{eN}\stackrel{>}{{}_\sim}  10^{-2}$, whereas  the process~(ii)
becomes   only   relevant   if   $B_{eN}$   is   unobservably   small,
e.g.~$B_{eN}\stackrel{<}{{}_\sim} 10^{-3}$.

\begin{figure}[t]
\begin{center}
{
\unitlength=1.3 pt
\SetScale{1.3}
\SetWidth{0.7}      
{} \qquad\allowbreak
\begin{picture}(79,65)(0,0)
\Text(13.0,49.0)[r]{$e^-$}
\ArrowLine(14.0,49.0)(31.0,49.0) 
\Text(39.0,50.0)[b]{$N$}
\Line(31.0,49.0)(48.0,49.0) 
\Text(66.0,57.0)[l]{$l^+$}
\ArrowLine(65.0,57.0)(48.0,49.0) 
\Text(66.0,41.0)[l]{$W^-$}
\Photon(65.0,41.0)(48.0,49.0){3.0}{2.0} 
\Text(27.0,41.0)[r]{$W$}
\Photon(31.0,33.0)(31.0,49.0){3.0}{2.0} 
\Text(13.0,33.0)[r]{$\gamma$}
\Photon(14.0,33.0)(31.0,33.0){3.0}{2.0} 
\Photon(31.0,33.0)(48.0,33.0){3.0}{2.0} 
\Text(66.0,25.0)[l]{$W^-$}
\Photon(65.0,25.0)(48.0,33.0){3.0}{2.0} 
\Text(39,0)[b] {\bf (a)}
\end{picture} \ 
{} \qquad\allowbreak
\begin{picture}(79,65)(0,0)
\Text(13.0,57.0)[r]{$e^-$}
\ArrowLine(14.0,57.0)(31.0,49.0) 
\Text(13.0,41.0)[r]{$\gamma$}
\Photon(14.0,41.0)(31.0,49.0){3.0}{2.0} 
\Text(39.0,53.0)[b]{$e^-$}
\ArrowLine(31.0,49.0)(48.0,49.0) 
\Text(66.0,57.0)[l]{$W^-$}
\Photon(65.0,57.0)(48.0,49.0){3.0}{2.0} 
\Text(47.0,41.0)[r]{$N$}
\Line(48.0,49.0)(48.0,33.0) 
\Text(66.0,41.0)[l]{$l^+$}
\ArrowLine(65.0,41.0)(48.0,33.0) 
\Text(66.0,25.0)[l]{$W^-$}
\Photon(65.0,25.0)(48.0,33.0){3.0}{2.0} 
\Text(39,0)[b] {\bf (b)}
\end{picture} \ 
{} \qquad\allowbreak
\begin{picture}(79,65)(0,0)
\Text(13.0,57.0)[r]{$e^-$}
\ArrowLine(14.0,57.0)(48.0,57.0) 
\Text(66.0,57.0)[l]{$W^-$}
\Photon(65.0,57.0)(48.0,57.0){3.0}{2.0} 
\Text(47.0,49.0)[r]{$N$}
\Line(48.0,57.0)(48.0,41.0) 
\Text(66.0,41.0)[l]{$W^-$}
\Photon(65.0,41.0)(48.0,41.0){3.0}{2.0} 
\Text(44.0,33.0)[r]{$l$}
\ArrowLine(48.0,25.0)(48.0,41.0) 
\Text(13.0,25.0)[r]{$\gamma$}
\Photon(14.0,25.0)(48.0,25.0){3.0}{4.0} 
\Text(66.0,25.0)[l]{$l^+$}
\ArrowLine(65.0,25.0)(48.0,25.0) 
\Text(39,0)[b] {\bf (c)}
\end{picture} \ 
{} \qquad\allowbreak
\begin{picture}(79,65)(0,0)
\Text(13.0,57.0)[r]{$e^-$}
\ArrowLine(14.0,57.0)(48.0,57.0) 
\Text(66.0,57.0)[l]{$W^-$}
\Photon(65.0,57.0)(48.0,57.0){3.0}{2.0} 
\Text(47.0,49.0)[r]{$N$}
\Line(48.0,57.0)(48.0,41.0) 
\Text(66.0,41.0)[l]{$l^+$}
\ArrowLine(65.0,41.0)(48.0,41.0) 
\Text(44.0,33.0)[r]{$W$}
\Photon(48.0,25.0)(48.0,41.0){3.0}{2.0} 
\Text(13.0,25.0)[r]{$\gamma$}
\Photon(14.0,25.0)(48.0,25.0){3.0}{4.0} 
\Text(66.0,25.0)[l]{$W^-$}
\Photon(65.0,25.0)(48.0,25.0){3.0}{2.0} 
\Text(39,0)[b] {\bf (d)}
\end{picture} \ 
}
\end{center}
\caption{\it Feynman diagrams for the process $e^-\gamma\rightarrow
W^-W^-l^+$.}
\label{A}
\end{figure}
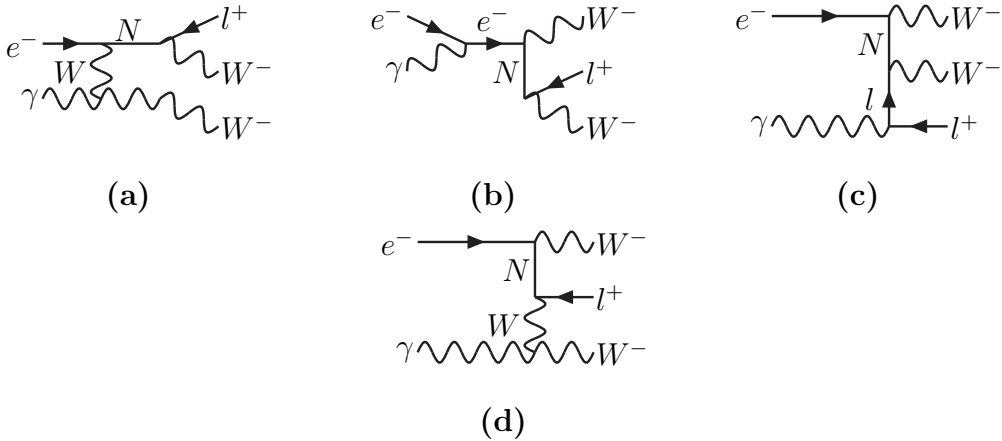

\subsection{\boldmath $B_{eN}\ne 0$}

We first consider  the process $e^-\gamma\rightarrow W^-W^-l^+$, which
becomes the dominant signal for $B_{eN}\stackrel{>}{{}_\sim} 10^{-2}$.
The  Feynman   diagrams  pertinent  to  this  process   are  shown  in
Fig.~\ref{A}~\footnote{All     diagrams,    including    Figs.~\ref{B}
and~\ref{C}, are produced with {\tt Axodraw}~\cite{Vermaseren:1994je}.}.  We
improve       upon      an       earlier      study       of      this
reaction~\cite{Peressutti:2001ms},   by   carefully  considering   the
contributing SM background (see our discussion in Section 3.3).

\begin{figure}[t]
\includegraphics{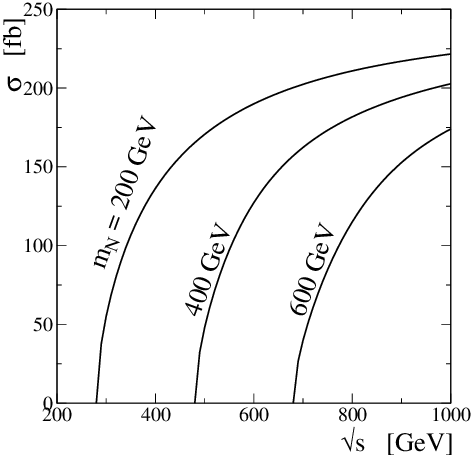}
\includegraphics{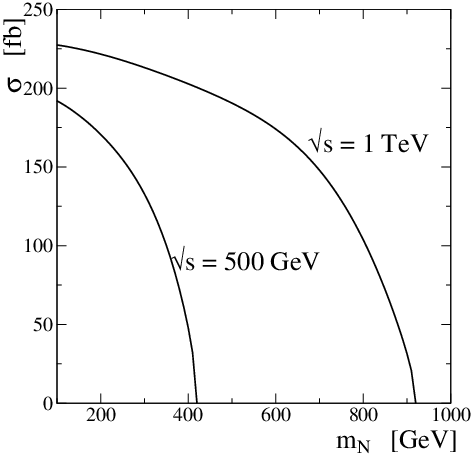}
\caption{\it  Numerical   estimates  of  the   cross  section  $\sigma
(e^-\gamma\rightarrow W^-N)$  as functions of  $\sqrt{s}$ (left frame)
and $m_N$ (right frame), with $B_{eN}=0.07$.}
\label{E}
\end{figure}

The  reaction  $e^-\gamma\rightarrow W^-W^-l^+$  is  dominated by  the
graphs (a) and (b) of Fig.~\ref{A}, where the heavy Majorana neutrinos
occur in the $s$-channel.  Considering therefore the $2 \rightarrow 2$
sub-process $e^-\gamma\rightarrow W^-N$,  Fig.~\ref{E} gives the cross
section   $\sigma  (e^-\gamma\rightarrow   W^-N)$   as  functions   of
$\sqrt{s}$  and  $m_N$.    The  differential  cross  sections  related
to~Fig.~\ref{E}, as  well as to  Figs.~\ref{H} and \ref{F},  have been
calculated using the  {\tt FeynCalc} package \cite{Mertig:1990an}.  We
have   also    checked   that    our   results   are    in   agreement
with~\cite{Peressutti:2001ms}.

Approximate    values    for    the    cross   section    $\sigma    (
e^-\gamma\rightarrow W^-W^-l^+)  $ can be obtained  by multiplying the
$y$-axis  values  from Fig.~\ref{E}  with  the  branching fraction  $B
(N\rightarrow  W^-l^+)$.  However, as  well as  ignoring contributions
from   off-shell  heavy  neutrinos,   this  approximation   makes  the
imposition  of kinematical  cuts on  the three-momentum  of  the final
charged  lepton  rather  difficult.    Therefore,  in  Section  4,  we
calculate the LNV signatures by considering the complete $2\rightarrow
3$ process,  as it is  described by the  full set of  Feynman diagrams
depicted  in  Fig.~\ref{A}.   To  this  end,  we  have  extended  {\tt
CompHEP}~\cite{Boos:2004kh}   to  include   heavy   Majorana  neutrino
interactions.

\subsection{\boldmath $B_{eN}= 0$}

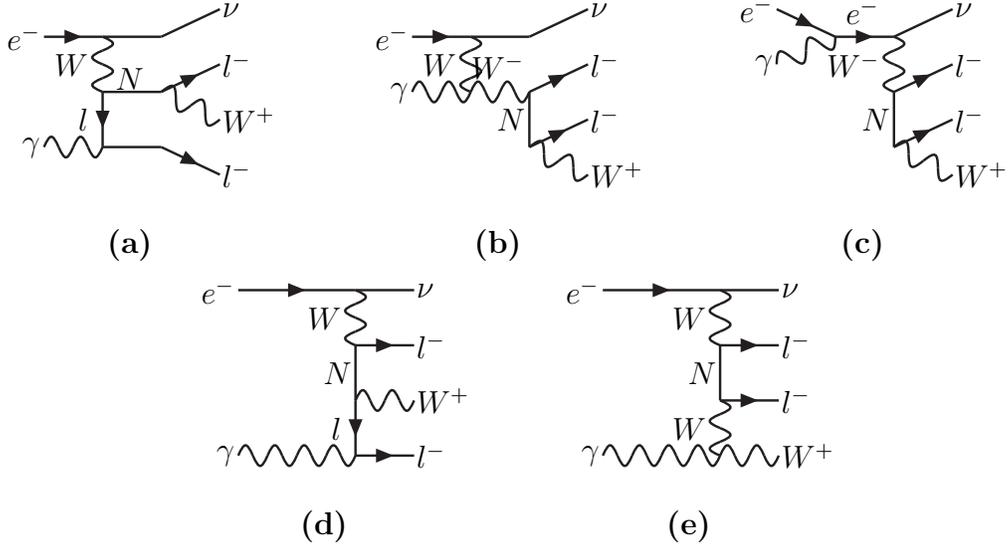
\begin{figure}[t]
\begin{center}
{
\unitlength=1.3 pt
\SetScale{1.3}
\SetWidth{0.7}      
{} \qquad\allowbreak
\begin{picture}(80,100)(0,0)
\Text(13.0,65.0)[r]{$e^-$}
\ArrowLine(14.0,65.0)(31.0,65.0) 
\Line(31.0,65.0)(48.0,65.0) 
\Text(66.0,73.0)[l]{$\nu$}
\Line(48.0,65.0)(65.0,73.0) 
\Text(27.0,57.0)[r]{$W$}
\Photon(31.0,49.0)(31.0,65.0){3.0}{2.0} 
\Text(39.0,50.0)[b]{$N$}
\Line(31.0,49.0)(48.0,49.0) 
\Text(66.0,57.0)[l]{$l^-$}
\ArrowLine(48.0,49.0)(65.0,57.0) 
\Text(66.0,41.0)[l]{$W^+$}
\Photon(48.0,49.0)(65.0,41.0){3.0}{2.0} 
\Text(27.0,41.0)[r]{$l$}
\ArrowLine(31.0,49.0)(31.0,33.0) 
\Text(13.0,33.0)[r]{$\gamma$}
\Photon(14.0,33.0)(31.0,33.0){3.0}{2.0} 
\Line(31.0,33.0)(48.0,33.0) 
\Text(66.0,25.0)[l]{$l^-$}
\ArrowLine(48.0,33.0)(65.0,25.0) 
\Text(39,0)[b] {\bf (a)}
\end{picture} \ 
{} \qquad\allowbreak
\begin{picture}(79,81)(0,0)
\Text(13.0,65.0)[r]{$e^-$}
\ArrowLine(14.0,65.0)(31.0,65.0) 
\Line(31.0,65.0)(48.0,65.0) 
\Text(66.0,73.0)[l]{$\nu$}
\Line(48.0,65.0)(65.0,73.0) 
\Text(27.0,57.0)[r]{$W$}
\Photon(31.0,49.0)(31.0,65.0){3.0}{2.0} 
\Text(13.0,49.0)[r]{$\gamma$}
\Photon(14.0,49.0)(31.0,49.0){3.0}{2.0} 
\Text(39.0,53.0)[b]{$W^-$}
\Photon(48.0,49.0)(31.0,49.0){3.0}{2.0} 
\Text(66.0,57.0)[l]{$l^-$}
\ArrowLine(48.0,49.0)(65.0,57.0) 
\Text(47.0,41.0)[r]{$N$}
\Line(48.0,49.0)(48.0,33.0) 
\Text(66.0,41.0)[l]{$l^-$}
\ArrowLine(48.0,33.0)(65.0,41.0) 
\Text(66.0,25.0)[l]{$W^+$}
\Photon(48.0,33.0)(65.0,25.0){3.0}{2.0} 
\Text(39,0)[b] {\bf (b)}
\end{picture} \
{} \qquad\allowbreak
\begin{picture}(79,81)(0,0)
\Text(13.0,73.0)[r]{$e^-$}
\ArrowLine(14.0,73.0)(31.0,65.0) 
\Text(13.0,57.0)[r]{$\gamma$}
\Photon(14.0,57.0)(31.0,65.0){3.0}{2.0} 
\Text(39.0,69.0)[b]{$e^-$}
\ArrowLine(31.0,65.0)(48.0,65.0) 
\Text(66.0,73.0)[l]{$\nu$}
\Line(48.0,65.0)(65.0,73.0) 
\Text(44.0,57.0)[r]{$W^-$}
\Photon(48.0,49.0)(48.0,65.0){3.0}{2.0} 
\Text(66.0,57.0)[l]{$l^-$}
\ArrowLine(48.0,49.0)(65.0,57.0) 
\Text(47.0,41.0)[r]{$N$}
\Line(48.0,49.0)(48.0,33.0) 
\Text(66.0,41.0)[l]{$l^-$}
\ArrowLine(48.0,33.0)(65.0,41.0) 
\Text(66.0,25.0)[l]{$W^+$}
\Photon(48.0,33.0)(65.0,25.0){3.0}{2.0} 
\Text(39,0)[b] {\bf (c)}
\end{picture} \  
{} \qquad\allowbreak
\begin{picture}(79,81)(0,0)
\Text(13.0,73.0)[r]{$e^-$}
\ArrowLine(14.0,73.0)(48.0,73.0) 
\Text(66.0,73.0)[l]{$\nu$}
\Line(48.0,73.0)(65.0,73.0) 
\Text(44.0,65.0)[r]{$W$}
\Photon(48.0,57.0)(48.0,73.0){3.0}{2.0} 
\Text(66.0,57.0)[l]{$l^-$}
\ArrowLine(48.0,57.0)(65.0,57.0) 
\Text(47.0,49.0)[r]{$N$}
\Line(48.0,57.0)(48.0,41.0) 
\Text(66.0,41.0)[l]{$W^+$}
\Photon(48.0,41.0)(65.0,41.0){3.0}{2.0} 
\Text(44.0,33.0)[r]{$l$}
\ArrowLine(48.0,41.0)(48.0,25.0) 
\Text(13.0,25.0)[r]{$\gamma$}
\Photon(14.0,25.0)(48.0,25.0){3.0}{4.0} 
\Text(66.0,25.0)[l]{$l^-$}
\ArrowLine(48.0,25.0)(65.0,25.0) 
\Text(39,0)[b] {\bf (d)}
\end{picture} \ 
{} \qquad\allowbreak
\begin{picture}(79,81)(0,0)
\Text(13.0,73.0)[r]{$e^-$}
\ArrowLine(14.0,73.0)(48.0,73.0) 
\Text(66.0,73.0)[l]{$\nu$}
\Line(48.0,73.0)(65.0,73.0) 
\Text(44.0,65.0)[r]{$W$}
\Photon(48.0,57.0)(48.0,73.0){3.0}{2.0} 
\Text(66.0,57.0)[l]{$l^-$}
\ArrowLine(48.0,57.0)(65.0,57.0) 
\Text(47.0,49.0)[r]{$N$}
\Line(48.0,57.0)(48.0,41.0) 
\Text(66.0,41.0)[l]{$l^-$}
\ArrowLine(48.0,41.0)(65.0,41.0) 
\Text(44.0,33.0)[r]{$W$}
\Photon(48.0,41.0)(48.0,25.0){3.0}{2.0}
\Text(13.0,25.0)[r]{$\gamma$}
\Photon(14.0,25.0)(48.0,25.0){3.0}{4.0}
\Text(66.0,25.0)[l]{$W^+$}
\Photon(48.0,25.0)(65.0,25.0){3.0}{2.0} 
\Text(39,0)[b] {\bf (e)}
\end{picture} \ 
}
\end{center}
\caption{\it Feynman diagrams for the process $e^-\gamma \rightarrow
W^+l^-l^-\nu$, with $B_{eN}=0$.}
\label{B}
\end{figure}

If the  coupling of the heavy  neutrino $N$ to the  electron is either
zero or very small, then the dominant process is $e^-\gamma\rightarrow
W^+l^-l^-\nu$,  where  $l=\mu,\tau$.  The  Feynman  diagrams for  this
process are shown in Fig.~\ref{B}.   Although the cross section of the
new  reaction is  much smaller  in this  case, so  is  the background.
Hence,  the predicted  LNV  signal could  still  be within  observable
reach. This is a major benefit compared to an $e^+e^-$ collider, which
requires  sizeable heavy  neutrino couplings  to the  electron  for an
observable signal  \footnote{There may exist other  processes as well,
such as  $e^-e^+ \to Z^* \to NN$  and $e^+ e^- \to  N \nu\nu\nu$.  The
first cross section is suppressed  by the small $Z$-coupling to a pair
of  heavy  neutrinos,  i.e.~it  is  ${\cal  O}(C^2_{NN})  =  {\cal  O}
(|B_{lN}|^4)$.   The   second  reaction  is   sub-dominant  ${\cal  O}
(\alpha^2_w  |B_{lN}|^2)$   due  to  the   additional  gauge  coupling
constants  involved.   Moreover, the  cross  section  of the  possible
reaction $e^-e^- \to W^- W^-$ is suppressed of order $|B_{eN}|^4$, and
severely constrained by the current $0\nu\beta\beta$-decay data.}.  To
the best of  our knowledge, this is a novel  possibility which has not
been investigated before.

\begin{figure}[t]
  \vskip1cm
\includegraphics{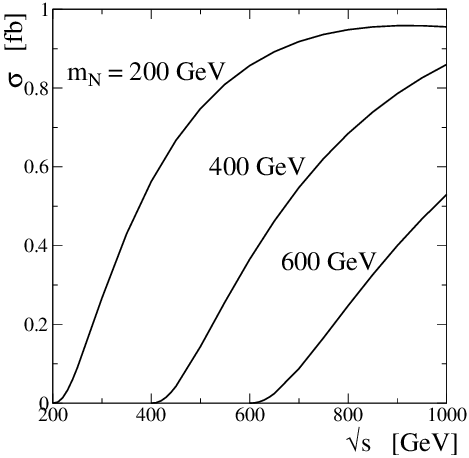}
\includegraphics{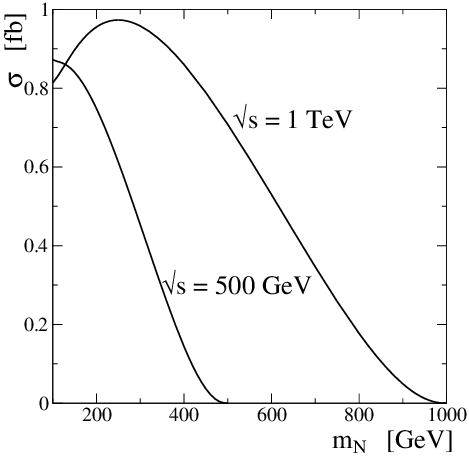}
\caption{\it  Numerical  estimates  of  the cross  section  $\sigma  (
e^-\gamma\rightarrow  N\mu^-\nu)$ versus  $\sqrt{s}$ (left  frame) and
$m_N$ (right frame), where $B_{eN}=0$, $B_{\mu N}=0.1$ and an infrared
angle cut $-0.99\le \cos\theta_{e^-\mu^-}\le0.99$ is used.}
\label{H}
\end{figure}

The  cross  section  for   the  LNV  reaction  $e^-\gamma  \rightarrow
W^+l^-l^-\nu$ is dominated by the resonant $s$-channel exchange graphs
(a)--(c) of the heavy Majorana neutrino $N$, as shown in Fig.~\ref{B}.
Thus, we  present in Fig.~\ref{H} the heavy  neutrino production cross
section  $\sigma (  e^-\gamma\rightarrow N\mu^-\nu)$  as  functions of
$\sqrt{s}$  and  $m_N$.   The  4-dimensional  phase-space  integration
involved in  the $2 \rightarrow  3$ process was  performed numerically
using  {\tt Bases}~\cite{Kawabata:1995th}.   The cross  section values
obtained for this process are about 2 orders of magnitude smaller than
those presented in Fig.~\ref{E}, for the reaction $e^-\gamma\to W^-N$.
An   approximate   estimate    of   $\sigma   (e^-\gamma   \rightarrow
W^+\mu^-\mu^-\nu)$ may  be obtained by  $\sigma ( e^-\gamma\rightarrow
N\mu^-\nu)  \times  B(N\rightarrow W^+\mu^-)$.   As  was mentioned  in
Section  3.1, this  method has  certain limitations  when differential
cross sections  are considered.   For this reason,  the LNV  signal in
Section 4 is computed using the  full set of the diagrams displayed in
Fig.~\ref{B} and  our extended version of {\tt  CompHEP} that includes
heavy Majorana neutrino interactions.

\subsection{The Standard Model Background}

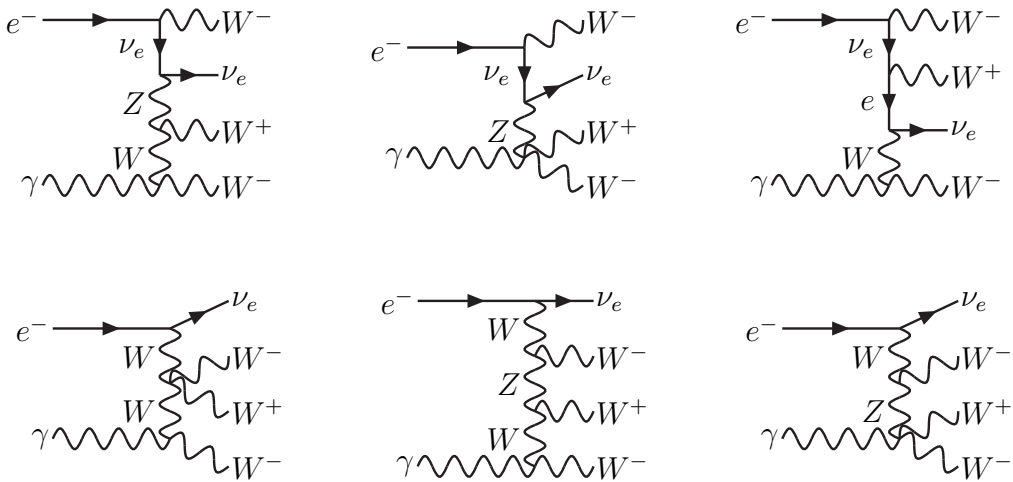
\begin{figure}[t]
\begin{center}
{
\unitlength=1.3 pt
\SetScale{1.3}
\SetWidth{0.7}      
{} \qquad\allowbreak
\begin{picture}(79,81)(0,0)
\Text(13.0,73.0)[r]{$e^-$}
\ArrowLine(14.0,73.0)(48.0,73.0) 
\Text(66.0,73.0)[l]{$W^-$}
\Photon(65.0,73.0)(48.0,73.0){3.0}{2.0} 
\Text(44.0,65.0)[r]{$\nu_e$}
\ArrowLine(48.0,73.0)(48.0,57.0) 
\Text(66.0,57.0)[l]{$\nu_e$}
\ArrowLine(48.0,57.0)(65.0,57.0) 
\Text(44.0,49.0)[r]{$Z$}
\Photon(48.0,57.0)(48.0,41.0){3.0}{2.0} 
\Text(66.0,41.0)[l]{$W^+$}
\Photon(65.0,41.0)(48.0,41.0){3.0}{2.0} 
\Text(44.0,33.0)[r]{$W$}
\Photon(48.0,41.0)(48.0,25.0){3.0}{2.0} 
\Text(13.0,25.0)[r]{$\gamma$}
\Photon(14.0,25.0)(48.0,25.0){3.0}{4.0} 
\Text(66.0,25.0)[l]{$W^-$}
\Photon(48.0,25.0)(65.0,25.0){3.0}{2.0} 
\end{picture} \ 
{} \qquad\allowbreak
\begin{picture}(79,81)(0,0)
\Text(13.0,65.0)[r]{$e^-$}
\ArrowLine(14.0,65.0)(48.0,65.0) 
\Text(66.0,73.0)[l]{$W^-$}
\Photon(65.0,73.0)(48.0,65.0){3.0}{2.0} 
\Text(44.0,57.0)[r]{$\nu_e$}
\ArrowLine(48.0,65.0)(48.0,49.0) 
\Text(66.0,57.0)[l]{$\nu_e$}
\ArrowLine(48.0,49.0)(65.0,57.0) 
\Text(44.0,41.0)[r]{$Z$}
\Photon(48.0,49.0)(48.0,33.0){3.0}{2.0}
\Text(13.0,33.0)[r]{$\gamma$}
\Photon(14.0,33.0)(48.0,33.0){3.0}{4.0} 
\Text(66.0,41.0)[l]{$W^+$}
\Photon(48.0,33.0)(65.0,41.0){3.0}{2.0} 
\Text(66.0,25.0)[l]{$W^-$}
\Photon(65.0,25.0)(48.0,33.0){3.0}{2.0} 
\end{picture} \ 
{} \qquad\allowbreak
\begin{picture}(79,81)(0,0)
\Text(13.0,73.0)[r]{$e^-$}
\ArrowLine(14.0,73.0)(48.0,73.0) 
\Text(66.0,73.0)[l]{$W^-$}
\Photon(65.0,73.0)(48.0,73.0){3.0}{2.0} 
\Text(44.0,65.0)[r]{$\nu_e$}
\ArrowLine(48.0,73.0)(48.0,57.0) 
\Text(66.0,57.0)[l]{$W^+$}
\Photon(48.0,57.0)(65.0,57.0){3.0}{2.0} 
\Text(44.0,49.0)[r]{$e$}
\ArrowLine(48.0,57.0)(48.0,41.0) 
\Text(66.0,41.0)[l]{$\nu_e$}
\ArrowLine(48.0,41.0)(65.0,41.0) 
\Text(44.0,33.0)[r]{$W$}
\Photon(48.0,25.0)(48.0,41.0){3.0}{2.0} 
\Text(13.0,25.0)[r]{$\gamma$}
\Photon(14.0,25.0)(48.0,25.0){3.0}{4.0} 
\Text(66.0,25.0)[l]{$W^-$}
\Photon(65.0,25.0)(48.0,25.0){3.0}{2.0} 
\end{picture} \ 
{} \qquad\allowbreak
\begin{picture}(79,81)(0,0)
\Text(13.0,65.0)[r]{$e^-$}
\ArrowLine(14.0,65.0)(48.0,65.0) 
\Text(66.0,73.0)[l]{$\nu_e$}
\ArrowLine(48.0,65.0)(65.0,73.0) 
\Text(44.0,57.0)[r]{$W$}
\Photon(48.0,49.0)(48.0,65.0){3.0}{2.0} 
\Text(66.0,57.0)[l]{$W^-$}
\Photon(65.0,57.0)(48.0,49.0){3.0}{2.0} 
\Text(66.0,41.0)[l]{$W^+$}
\Photon(65.0,41.0)(48.0,49.0){3.0}{2.0} 
\Text(44.0,41.0)[r]{$W$}
\Photon(48.0,49.0)(48.0,33.0){3.0} {2.0}
\Text(13.0,33.0)[r]{$\gamma$}
\Photon(14.0,33.0)(48.0,33.0){3.0}{4.0} 
\Text(66.0,25.0)[l]{$W^-$}
\Photon(48.0,33.0)(65.0,25.0){3.0}{2.0} 
\end{picture} \ 
{} \qquad\allowbreak
\begin{picture}(79,81)(0,0)
\Text(13.0,73.0)[r]{$e^-$}
\ArrowLine(14.0,73.0)(48.0,73.0) 
\Text(66.0,73.0)[l]{$\nu_e$}
\ArrowLine(48.0,73.0)(65.0,73.0) 
\Text(44.0,65.0)[r]{$W$}
\Photon(48.0,57.0)(48.0,73.0){3.0}{2.0} 
\Text(66.0,57.0)[l]{$W^-$}
\Photon(65.0,57.0)(48.0,57.0){3.0} {2.0}
\Text(44.0,49.0)[r]{$Z$}
\Photon(48.0,57.0)(48.0,41.0){3.0}{2.0}
\Text(66.0,41.0)[l]{$W^+$}
\Photon(65.0,41.0)(48.0,41.0){3.0}{2.0} 
\Text(44.0,33.0)[r]{$W$}
\Photon(48.0,41.0)(48.0,25.0){3.0}{2.0} 
\Text(13.0,25.0)[r]{$\gamma$}
\Photon(14.0,25.0)(48.0,25.0){3.0}{4.0} 
\Text(66.0,25.0)[l]{$W^-$}
\Photon(48.0,25.0)(65.0,25.0){3.0} {2.0}
\end{picture} \ 
{} \qquad\allowbreak
\begin{picture}(79,81)(0,0)
\Text(13.0,65.0)[r]{$e^-$}
\ArrowLine(14.0,65.0)(48.0,65.0) 
\Text(66.0,73.0)[l]{$\nu_e$}
\ArrowLine(48.0,65.0)(65.0,73.0) 
\Text(44.0,57.0)[r]{$W$}
\Photon(48.0,49.0)(48.0,65.0){3.0}{2.0} 
\Text(66.0,57.0)[l]{$W^-$}
\Photon(65.0,57.0)(48.0,49.0){3.0}{2.0}
\Text(44.0,41.0)[r]{$Z$}
\Photon(48.0,49.0)(48.0,33.0){3.0}{2.0} 
\Text(13.0,33.0)[r]{$\gamma$}
\Photon(14.0,33.0)(48.0,33.0){3.0}{4.0} 
\Text(66.0,41.0)[l]{$W^+$}
\Photon(48.0,33.0)(65.0,41.0){3.0}{2.0} 
\Text(66.0,25.0)[l]{$W^-$}
\Photon(65.0,25.0)(48.0,33.0){3.0}{2.0} 
\end{picture} \ 
}
\end{center}
\caption{\it Typical dominant Feynman diagrams related to the SM
reaction $e^-\gamma\rightarrow W^-W^-W^+\nu_e$.}
\label{C}
\end{figure}

The LNV reactions  we have been considering are  strictly forbidden in
the SM. The contributing SM  background that may mimic the signal will
always  involve  additional  light  neutrinos.  Specifically,  the  SM
background  arises from  the resonant  production and  decay  of three
$W^\pm$  bosons, i.e.~$e^-\gamma\rightarrow  W^-W^-W^+\nu_e$.  Typical
dominant  graphs of  this $2\to  4$  scattering process  are shown  in
Fig.~\ref{C}.

The  background  to   the  first  LNV  reaction  $e^-\gamma\rightarrow
W^-W^-l^+$  originates  from   the  decay  $W^+\rightarrow  l^+\nu_l$.
Correspondingly,   the   background   to   the  second   LNV   process
$e^-\gamma\rightarrow W^+l^-l^-\nu$  results from the  leptonic decays
$W^-\rightarrow l^-\bar{\nu}_l$  for both of  the $W^-$ bosons  in the
final state.  To  compute the background process $e^-\gamma\rightarrow
W^-W^-W^+\nu_e$, we first  use {\tt CompHEP}~\cite{Boos:2004kh}, which
also  generates the  appropriate weighted  events, and  then  use {\tt
PYTHIA}  \cite{Sjostrand:2000wi}   interfaced  via  the   {\tt  CPyth}
program~\cite{Belyaev:2000wn}.  The last step is necessary in order to
properly describe the  Lorentz-boosted $W^\pm$-boson decay products on
which  appropriate kinematical  cuts were  placed.   These kinematical
cuts will  be presented in  detail in Section~4.  Although  our method
uses the branching fractions for the $W$ decays \cite{Eidelman:2004wy}
as an  input, it proves by  far more practical  than generating events
for the  complete set of $2\to  7$ background processes,  where a vast
number of off-resonant amplitudes give negligible contributions.

It  is  important  to  clarify  at  this  point  that  observation  of
$e^-\gamma \to W^+l^-l^-\nu$ does not constitute by itself a signature
for  LNV, even  if $l\ne  e$.  If  the heavy  neutrino couples  to the
electron as well as to the  muon or tau lepton, then the occurrence of
such a  reaction will only  signify lepton flavour violation.   If one
now  assumes that  $e^-\gamma\rightarrow W^-W^-l^+$  is  not observed,
there are  two possibilities  that could result  in an  observation of
$e^-\gamma\rightarrow W^+l^-l^-\nu$.   Either the heavy  neutrino is a
Majorana particle that does  not couple predominantly to the electron,
or it is a Dirac particle that does.  To distinguish between these two
possibilities, one  has to look  for $e^-\gamma\rightarrow W^-W^+l^-$,
where  both $W^\pm$  bosons decay  hadronically.  This  process cannot
occur,  unless the  heavy  neutrino  has a  relevant  coupling to  the
electron. For a heavy Dirac  neutrino, however, the latter will always
be larger than $e^-\gamma\rightarrow W^+l^-l^-\nu$.  Therefore, if the
reactions $e^-\gamma\to W^-W^\mp l^\pm$ were not detected, observation
of $e^-\gamma  \to W^+l^-l^-\nu$ would  be a clear manifestation  of a
LNV signature, which is mediated by a heavy Majorana neutrino.

\setcounter{equation}{0}
\section{Numerical Results}

There  are many  theoretical parameters  in the  SM  with right-handed
neutrinos  that  could  vary  independently, such  as  heavy  Majorana
neutrino masses and couplings.  In addition, the machine parameters of
any future $e^-\gamma$ collider  are still under discussion.  Since it
would be  counter-productive to  explore all possible  situations that
may occur, we will,  instead, analyze two representative scenarios for
each  of  the   processes:  (i)~$e^-\gamma\rightarrow  W^-W^-e^+$  and
(ii)~$e^-\gamma\rightarrow W^+\mu^-\mu^-\nu$.

In  our  analysis,  we   take  $\vert  {B_{lN}\vert}^2  \approx  \vert
{C_{\nu_i  N}\vert}^2$,  i.e.~we assume  that  only  one flavour  will
couple  predominantly  to  the   heavy  Majorana  neutrino  $N$.   The
Higgs-boson   mass  value  $M_H=120\,$GeV   is  used   throughout  our
estimates, but our results depend  only very weakly on $M_H$. Finally,
we use the global kinematical cut,
\begin{equation}
-0.99\ \le\ \cos\,\theta_{e^-l^\pm}\ \le\ 0.99\; ,
\label{7}
\end{equation}
for  the angle $\theta_{e^-l^\pm}$  between the  $e^-$ in  the initial
state  and  the  final-state  charged  leptons  $l^\pm$.   The  global
cut~(\ref{7})  was used to  ensure that  the produced  charged leptons
$l^\pm$ are detected.

\subsection{\boldmath $e^-\gamma\rightarrow W^-W^-e^+$}

We  will consider  the simplest  case where $B_{eN}$  is sizeable, but
$B_{\mu N}=B_{\tau N}=0$. We analyze two scenarios:

\noindent
{\em Scenario} (1):
$\qquad \sqrt{s}\ =\ 500~{\rm GeV}\,,\qquad m_N\ =\ 200~{\rm GeV}\,,\qquad
B_{eN}\ =\ 0.07\,,$\\[2mm]
predicting $\Gamma_N\ =\ 0.04~{\rm GeV}$ and 
$B(N\rightarrow W^-e^+)\ \approx\ 0.3$,\\[2mm] 
and

\noindent
{\em Scenario} (2): $\qquad \sqrt{s}\ =\ 1~{\rm TeV}\,,\qquad m_N\ =\
400~{\rm GeV}\,, \qquad B_{eN}\ =\ 0.07\; ,$\\[2mm]
predicting $\Gamma_N\ =\
0.4~{\rm GeV}$ and $B (N\rightarrow W^-e^+)\ \approx\ 0.25$.

\begin{figure}[t]
\begin{center}
\includegraphics{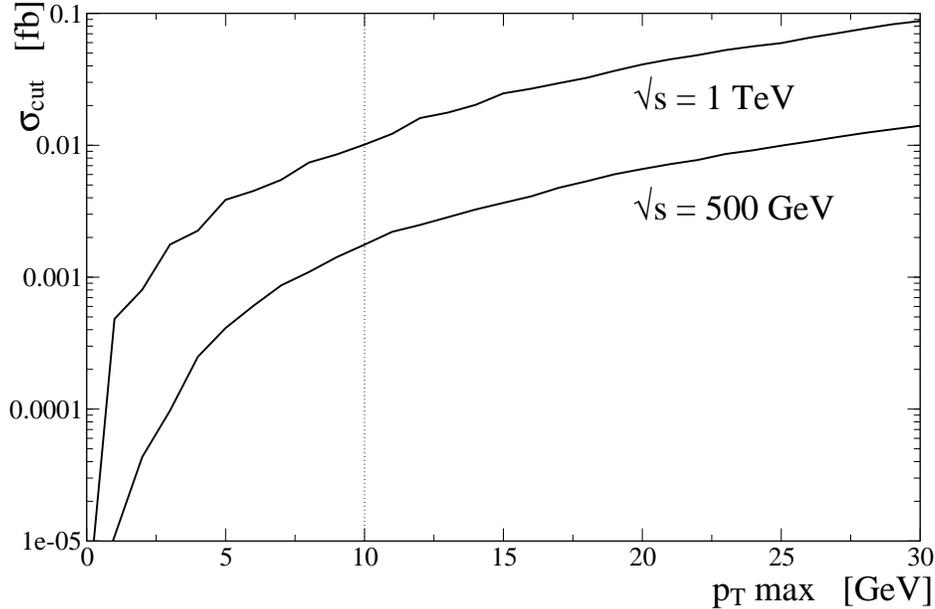}
\caption{\it   The  dependence   of  the   background   cross  section
$\sigma_{\rm   cut}  (e^-\gamma\rightarrow  W^-W^-e^+\nu_e\nu_e)\times
B(W^-W^-\rightarrow {\rm hadrons})$ on  the $p^{\rm max}_T$ cut, where
$\sigma_{cut}=\int_0^{p^{\rm   max}_T}  dp_T\,  \frac{d\sigma}{dp_T}$.
The dotted line shows the value of cut used in Table 1.}
\label{D}
\end{center}
\end{figure}

If the two  $W^-$ bosons decay hadronically,  the LNV signal will have
no   missing momentum,  whilst  the    SM background $e^-\gamma    \to
W^-W^-e^+\nu_e\nu_e$  has  two light     neutrinos  that will   escape
detection. Making use of this fact,  one may apply the kinematical cut
on the missing transverse momentum $p_T$:
\begin{equation}
p_T\ \le\  p^{\rm max}_T\ =\ 10~{\rm GeV}\; .
\end{equation}
This  has  no  effect  on  the  signal,  but  reduces  the  background
significantly. The value 10~GeV has been chosen simply as a reasonable
limit of the detector resolution \cite{Badelek:2001xb}. Figure~\ref{D}
shows  the  dependence of  the  background  on  $p^{\rm max}_T$.   The
inaccuracies for  low values of  $p^{\rm max}_T$ that are  apparent in
Fig.~\ref{D}  should be  regarded as  numerical artifacts  due  to the
small  fraction  of  events  that  have  passed  the  $p^{\rm  max}_T$
selection criterion.

In Table~1  we present comparative numerical values  of cross sections
for the signal  and the background in Scenarios (1)  and (2), with and
without  the $p_T$  cut.   Our numerical  estimates  also include  the
branching fraction  for the  two $W^-$ bosons  to decay  into hadrons.
Table~1 shows  that the  background can be  drastically reduced  to an
almost unobservable level.  For a  given scenario, this enables one to
place sensitivity limits to the mixing factor $B_{eN}$.  Since the LNV
cross sections exhibited in Table~1 are roughly equal in Scenarios~(1)
and~(2), the limits  on $|B_{eN}|$ will be very  similar.  Assuming an
integrated   luminosity  of  $100\,{\rm   fb}^{-1}$,  one   must  have
$|B_{eN}|\ge 4.6\times 10^{-3}$ to  obtain at least 10 background-free
events.  If  no signal  is observed, this  will place the  upper limit
$|B_{eN}|\le  2.7\times   10^{-3}$  at  90\%~CL.    Observe  that  the
sensitivity of an $e^-\gamma$ collider to $B_{eN}$ is better than that
of an $e^+e^-$ linear  collider~\cite{delAguila:2005mf} due to the far
smaller SM background.

\begin{table}[t]
\begin{center}
\begin{tabular}{|c||c|c|} \hline
Process & Scenario (1) & Scenario (2) \\ \hline \hline
Signal (without cut) & 22.9 & 23.3  \\
Background (without cut) & 0.11 & 1.54 \\ \hline
Signal (with cut) & 22.9 & 23.3  \\
Background (with cut) & 0.002 & 0.01 \\ \hline
\end{tabular}
\end{center}
\caption{\it    Cross    sections   (in    fb)    for   the    process
$e^-\gamma\rightarrow W^-W^-e^+$ and its SM background.  The branching
fraction B($W^-W^-\rightarrow$ hadrons) is included.}
\end{table}

\subsection{\boldmath $e^-\gamma\rightarrow W^+\mu^-\mu^-\nu$}

As was  mentioned  in Section~3,  this process becomes  relevant, only
when the mixing factor  $|B_{eN}| \stackrel{<}{{}_\sim} 10^{-3}$.   We
therefore  consider    cases  where  $B_{\mu  N}$   is   sizeable, but
$B_{eN}=B_{\tau N}=0$.   As    before, we analyze the   following  two
scenarios:

\noindent
{\em Scenario} (3): $\qquad 
\sqrt{s}\ =\ 500~{\rm GeV}\,,\qquad m_N\ =\ 200~{\rm GeV}\,,\qquad
B_{\mu N}=0.1\; ,$\\[2mm]
yielding $\Gamma_N\ =\ 0.08~{\rm GeV}$ and $B (N\rightarrow W^+\mu^-)\
\approx\ 0.3$,\\[2mm] and

\noindent
{\em Scenario} (4):
$\qquad
\sqrt{s}\ =\ 1~{\rm TeV}\,,\qquad m_N\ =\ 400~{\rm GeV}\,,\qquad 
B_{\mu N}\ =\ 0.1\; ,$\\[2mm]
yielding $\Gamma_N\ =\ 0.8~{\rm GeV}$ and 
$B (N\rightarrow W^+\mu^-)\ \approx\ 0.26$.

\begin{figure}[t]
\begin{center}
\includegraphics{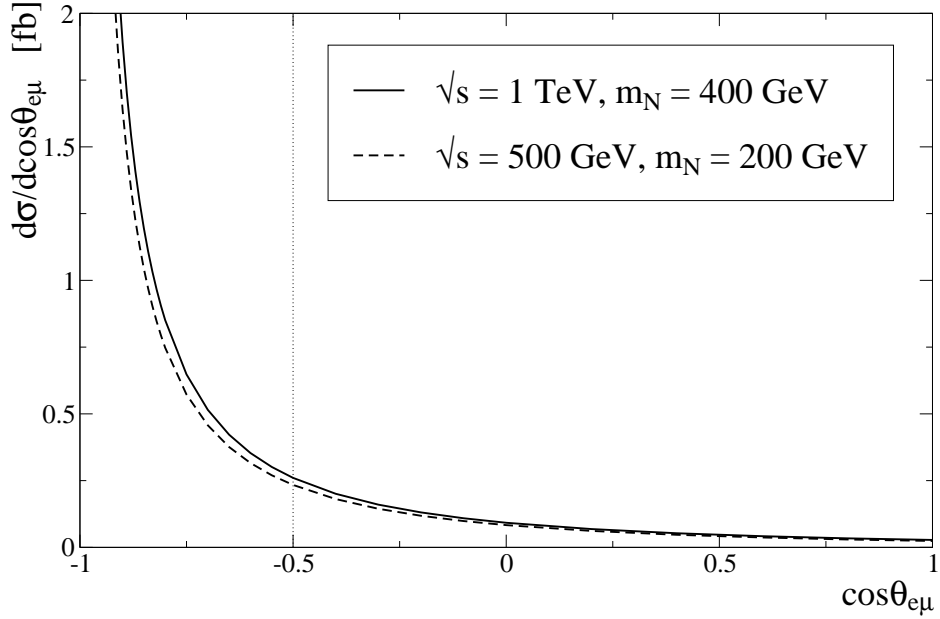}
\end{center}
\caption{\it    Differential   cross    section   for    the   process
$e^-\gamma\rightarrow  N\mu^-\nu$ ($B_{eN}=0$,  $B_{\mu  N}=0.1$). Our
cuts veto the region to the right of the dotted line.}
\label{F}
\end{figure}

\begin{figure}[t]
 \vskip1cm
\includegraphics{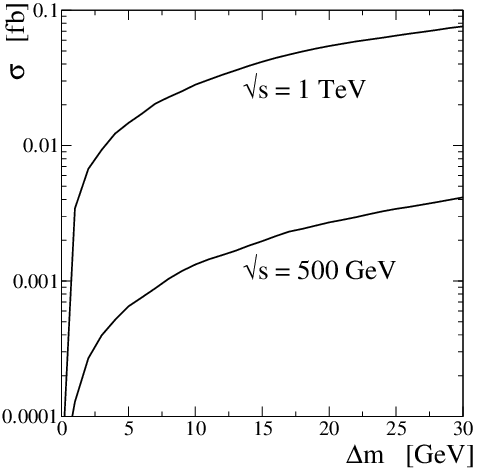}
\includegraphics{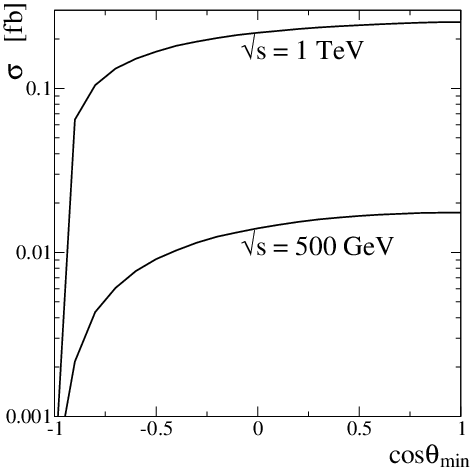}
\caption{\it The dependence of the background cross section
$\sigma(e^-\gamma\rightarrow
W^+\mu^-\mu^-\nu_e\bar\nu_\mu\bar\nu_\mu)\times B(W^+\rightarrow
{\rm hadrons})$ on the kinematical cuts $\Delta m$ (left panel) and
$\cos\theta_{\rm min}$ (right panel).}
\label{G}
\end{figure}

Since the heavy  Majorana neutrino will have a  small decay width, one
expects the invariant mass $m_{\mu^-W^+}$ of one of the muons with the
$W^+$ boson to  be very close to the mass of  the heavy neutrino.  The
other muon, which does not come from the decaying heavy neutrino, will
generally  have a  preference to  a  small scattering  angle from  the
direction of the incoming photon,  due to the infrared property of the
exchanged muon in the $t$-channel.  This last feature is also shown in
Fig.~\ref{F}.  In  view of all  the above reasons, the  following cuts
prove very effective:
\begin{eqnarray}
 \label{cut2}
m_{\mu^-W^+} &=& m_N\pm \Delta m\;,\qquad\qquad\, 
                             \mbox{for one of the muons,}\nonumber\\ 
\cos\theta_{e^-\mu^-} &\le& \cos\theta_{\rm min} = -0.5\;, \qquad 
\mbox{for the other muon.}
\end{eqnarray}
Here  $\Delta m =  10$~GeV is  again set  by the  detector resolution,
which  is much  greater than  the  actual heavy  neutrino decay  width
$\Gamma_N$.  Figure~\ref{G} shows how the background would be affected
by  different choices  of these  cuts. 

\begin{table}[b]
\begin{center}
\begin{tabular}{|c||c|c|} \hline
Process & Scenario (3) & Scenario (4) \\ \hline \hline
Signal (without cuts) & 0.15 & 0.15  \\
Background (without cuts) & 0.02 & 0.25 \\ \hline
Signal (with cuts) & 0.13 & 0.12  \\
Background (with cuts) & 0.0002 & 0.004 \\ \hline
\end{tabular}
\end{center}
\caption{\it Cross sections (in fb) for $e^-\gamma\rightarrow
W^+\mu^-\mu^-\nu$ and its SM background, where B($W^+\rightarrow$
hadrons) is included.}
\end{table}

In Table~2 we summarize our results for the signal and the background,
before   and  after  the   kinematical  cuts~(\ref{cut2})   have  been
implemented.   In  particular,  it  can  be  seen  that  the  selected
kinematical  cuts  were  very  effective  to  drastically  reduce  the
background by 2 orders of  magnitude, without harming much the signal.
The  limits that  can be  derived for  Scenarios~(3) and~(4)  are very
similar.  To observe 10 background-free events, one would need a $W\mu
N$-coupling  $|B_{\mu N}|  \ge  9.0\times 10^{-2}$,  assuming a  total
integrated luminosity of 100~fb$^{-1}$.  The absence of any LNV signal
would put the  upper limit $|B_{\mu N}| \le 0.05$  at the 90\%~CL.  It
is important to remark that  the sensitivity limits that can be placed
on $|B_{\mu N}|$  at an $e^-\gamma$ collider will  be much higher than
the  previously  considered  studies  at $e^+e^-$  and  $\gamma\gamma$
colliders~\cite{delAguila:2005mf,Peressutti:2002nf}.

\bigskip\bigskip

\section{Conclusions}

We    have    analyzed    the   phenomenological    consequences    of
electroweak-scale heavy Majorana neutrinos at an $e^-\gamma$ collider.
We  have found  that  heavy  Majorana neutrinos,  with  masses $m_N  =
100$--400~GeV  and couplings $B_{eN}\sim  10^{-2}$ and  $B_{\mu N}\sim
10^{-1}$, will  become easily  observable at an  $e^-\gamma$ collider,
with  CMS energies  $\sqrt{s}  = 0.5$--1~TeV  and  a total  integrated
luminosity of 100~fb$^{-1}$.  Specifically, we have computed the cross
sections  for the LNV  reactions $e^-\gamma\rightarrow  W^-W^-e^+$ and
$e^-\gamma\rightarrow  W^+\mu^-\mu^-\nu$.  In  our  analysis, we  have
also  considered  the   contributing  SM  background,  which  involves
additional  light  neutrinos  in  the  final  state.   After  imposing
realistic missing $p_T$,  invariant mass and angle cuts,  we have been
able to  suppress the  SM background  by 2 orders  of magnitude  to an
unobservable level, without harming much the signal.

Observation of the reaction $e^-\gamma \to W^-W^-e^+$ would be a clear
signal for LNV and hence  for unravelling the possible Majorana nature
of heavy  neutrinos.  The process $e^-\gamma  \to W^+\mu^-\mu^-\nu$ is
also  a  clear  signal for  LNV  in  the  absence of  observations  of
$e^-\gamma\rightarrow  W^+W^-\mu^-$.  The  search  for heavy  Majorana
neutrinos  at  an  $e^-\gamma$   collider  adds  particular  value  to
analogous  searches  at an  $e^+e^-$  linear  collider.  Although  the
latter may be  built first, it would be more  difficult to discern the
Dirac or Majorana nature of possible heavy neutrinos. Another obstacle
facing $e^+e^-$ colliders is the large irreducible SM background which
has similar  kinematical characteristics  as the signal.   Instead, an
$e^-\gamma$ collider  provides a cleaner environment and  has a unique
potential  to  identify the  possible  Majorana  nature  of the  heavy
neutrinos, as  well as  probe lower coupling  strengths thanks  to its
highly reducible background.

The analysis presented in this  paper may contain some degree of model
dependence.  For  example, models that include new  heavy gauge bosons
may add new  relevant LNV interactions to the  Lagrangian.  As long as
these  additional contributions  are negligible,  our results  will be
robust.  In conclusion, an $e^-\gamma$ collider would provide valuable
information  about the  properties  of possible  heavy neutrinos  that
could    be    relevant    for   low-scale    resonant    leptogenesis
\cite{Pilaftsis:2005rv}, and  also shed light on the  structure of the
light neutrino masses and mixings.

\section*{Acknowledgments}
We thank  George Lafferty for pointing us  out Ref.~\cite{BaBar}.  The
work   of   SB   has    been   funded   by   the   PPARC   studentship
PPA/S/S/2003/03666. The work of JSL  has been supported in part by the
PPARC  research  grant:  PPA/G/O/2002/00471.  Finally,  AP  gratefully
acknowledges  the  partial  support  by  the  PPARC  research  grants:
PPA/G/O/2002/00471 and PP/C504286/1.

\end{document}